\documentclass[prl,twocolumn,showpacs,superscriptaddress]{revtex4}
\usepackage{amsfonts}
\usepackage{amsmath}
\usepackage{amssymb}
\usepackage{graphicx}

\begin{document}

\title{Exact
solutions of the Gross-Pitaevskii equation for stable vortex
modes}
\author{Lei Wu}
\affiliation{Institute of Nonlinear Physics, Zhejiang Normal
University, Jinhua, Zhejiang 321004, China}
\author{ Lu Li}
\affiliation{Institute of Theoretical Physics, Shanxi University, Taiyuan, Shanxi 030006,
China}
\author{Jie-Fang Zhang}
\affiliation{Institute of Nonlinear Physics, Zhejiang Normal University, Jinhua, Zhejiang
321004, China}
\author{Dumitru Mihalache}
\affiliation{Horia Hulubei National Institute for Physics and Nuclear Engineering,
Magurele-Bucharest 077125, Romania}
\author{Boris A. Malomed}
\affiliation{Department of Physical Electronics, Faculty of Engineering, Tel Aviv
University, Tel Aviv 69978, Israel}
\author{W. M. Liu}
\affiliation{Institute of Physics, Chinese Academy of Sciences, Beijing 100190, China}

\begin{abstract}

We construct exact solutions of the Gross-Pitaevskii equation for
solitary vortices, and approximate ones for fundamental solitons,
in 2D models of Bose-Einstein condensates with a spatially
modulated nonlinearity of either sign and a harmonic trapping
potential. The number of vortex-soliton (VS) modes is determined
by the discrete energy spectrum of a related linear
Schr\"{o}dinger equation. The VS families in the system with
the attractive and repulsive nonlinearity are mutually complementary. \emph{%
Stable} VSs with vorticity $S\geq 2$ and those corresponding to higher-order
radial states are reported for the first time, in the case of the attraction
and repulsion, respectively.
\end{abstract}

\maketitle

The realization of Bose-Einstein condensates (BECs) in dilute
quantum gases has drawn a great deal of interest to the dynamics
of nonlinear excitations in matter waves, such as dark \cite{dark}
and bright solitons \cite{bright}, vortices
\cite{vortex,BECvortex}, supervortices \cite{supervortex}, etc.
The work in this direction was strongly stimulated by many
similarities to solitons and vortices in optics \cite{optics}. In
the mean-field approximation, the BEC dynamics at ultra-low
temperatures is accurately described by the Gross-Pitaevskii
equation (GPE), the nonlinearity being determined by the $s$-wave
scattering length of interatomic collisions, which can be
controlled by means of the magnetic \cite{magnetic} or low-loss
optical \cite{optical} Feshbach-resonance technique, making
spatiotemporal \textquotedblleft management" of the local
nonlinearity
possible through the use of time-dependent and/or non-uniform fields \cite%
{optics}. In particular, a recently developed technique, which
allows one to \textquotedblleft paint" arbitrary one- and
two-dimensional (1D and 2D) average patterns by a rapidly moving
laser beam \cite{painting}, suggests new perspectives for the
application of the nonlinearity management based on the optical
Feshbach-resonance. A counterpart of the GPE, which is a basic
model in
nonlinear optics, is the nonlinear Schr\"{o}dinger equation (NLSE) \cite%
{optics}. In the latter case, the modulation of the local nonlinearity can
be implemented too--in particular, by means of indiffusion of a dopant
resonantly interacting with the light.

The search for exact solutions to the GPE/NLSE is an essential
direction in the studies of nonlinear matter and photonic waves.
In particular, stable soliton solutions are of great significance
to experiments and potential applications, as they precisely
predict conditions which should allow the creation of matter-wave
and optical solitons, including challenging situations which have
not yet been tackled in the experiment, such as 2D matter-wave
solitons and solitary vortices \cite{review}. In the 1D setting
with the spatially uniform nonlinearity, exact stable solutions
for bright and dark solitons have been constructed in special
cases, when the external potential is linear or quadratic
\cite{homogeneous}, and a specially designed inhomogeneous
nonlinearity may support bound states of an arbitrary number of
solitons \cite{beitia}. In the 2D geometry with the uniform
nonlinearity, exact delocalized solutions were constructed in the
presence of a periodic potential \cite{LDcarr2}. In the 2D
geometry, 1D bright and dark solitons are unstable, with the
bright ones suffering the breakup into a chain of collapsing
pulses, and the dark solitons splitting into vortex-antivortex
pairs \cite{dark,RDS}. Another approach to finding exact solitonic
states in both 1D and 2D settings (with the uniform nonlinearity),
along with the supporting potentials, was proposed in the form the
respective inverse problem \cite{Step}. As concerns vortices,
while delocalized ones have been created and extensively
investigated in self-repulsive BECs \cite{vortex}, and vortex
solitons (VSs) were created in photorefractive crystals equipped
with photonic lattices \cite{photorefr}, analytical solutions for
localized vortices have not been reported yet, and experimental
creation of stable VSs in self-attractive BEC and optical media
with the fundamental cubic nonlinearity remains a challenge to the
experiment.

In this work we demonstrate that exact VS solutions can be found in the 2D
GPE/NLSE, with a specially designed (and experimentally realizable) profile
of the radial modulation of the nonlinearity coefficient, of both attractive
and repulsive types. In fact, the search for localized solutions in 2D
models with diverse nonlinearity-modulation profiles has recently attracted
considerable attention \cite{2D}, although no exact solutions have been
reported. We consider only nonlinearity-modulation patterns which do not
change the sign, as zero-crossing would make the use of the FR problematic
\cite{magnetic,optical}.

The scaled form of the GPE/NLSE is $i{\psi }_{t}=-\nabla ^{2}\psi +g(r)|\psi
|^{2}\psi +V(r)\psi $, where $\psi $ is the BEC\ macroscopic wave function, $%
\nabla ^{2}$ the 2D Laplacian, and $g(r)$ is the nonlinearity coefficient
which, as well as external potential $V(r)$, is a function of radial
coordinate $r$. The same equation finds a straightforward implementation as
a spatial-domain model of the beam propagation in bulk media, with $t$, $%
\psi $, and $g(r)$ being the propagation distance, amplitude of the
electromagnetic field, and the local Kerr coefficient.

Assuming the stationary wave function as $\psi (r,\theta ,t)=\phi (r)\exp
(iS\theta -i\mu t)$, where $\theta $ is the azimuthal angle, $S$ an integer
vorticity, and $\mu $ the chemical potential (or propagation constant in
optics), leads to the equation for the real stationary wave function $\phi
(r)$: $\mu \phi =-\phi ^{\prime \prime }-r^{-1}\phi ^{\prime }+g(r)\phi ^{3}+%
\left[ {S^{2}r^{-2}+}V(r)\right] \phi $. For $S\neq 0$, $\phi (r)$ should
vary as $r^{|S|}$ at $r\rightarrow 0$, which is replaced by $\phi ^{\prime
}(r=0)=0$ for $S=0$. The localization requires $\phi (r=\infty )=0$.

Defining $\phi (r)\equiv \rho (r)U(R(r))$, $g(r)\equiv g_{0} r^{-2}\rho^{-6}
(r)$, with $R(r)\equiv \int_{0}^{r}s^{-1} \rho ^{-2}(s)ds $, one can find
that $\rho (r)$ and $U(R)$ obey the following equations:
\begin{align}
\rho ^{\prime \prime }+r^{-1}\rho ^{\prime }+\left( \mu
-V(r)-S^{2}r^{-2}\right) \rho & =Er^{-2}\rho ^{-3},  \label{evolution} \\
-d^{2}U/dR^{2}+g_{0}U^{3}& =EU,  \label{final}
\end{align}%
where $E$ and $g_{0}$ are constants. The reduction of the GPE/NLSE to Eq. (%
\ref{final}) helps one to find exact solutions, as the latter equation is
solvable in terms of elliptic functions. Then, if a solution to Eq. (\ref%
{evolution}) is known, one can construct exact solutions to the underlying
GPE/NLSE, while the nonlinearity-modulation profile admitting exact
solutions is determined by $\rho (r)$. Physical solutions impose
restrictions on $\rho $: from expressions for $R(r)$ and $g(r)$ it follows
that $\rho $ cannot change its sign; further, it must behave as $r^{-a}$
with $a\geq 1/3$ at $r\rightarrow 0$, and diverge ($\rho \rightarrow \infty $%
) at $r\rightarrow \infty $, so that the nonlinearity strength is bounded
and the integration in $R(r)$ converges.

\begin{figure}[t]
\includegraphics[width=7cm,height=4cm]{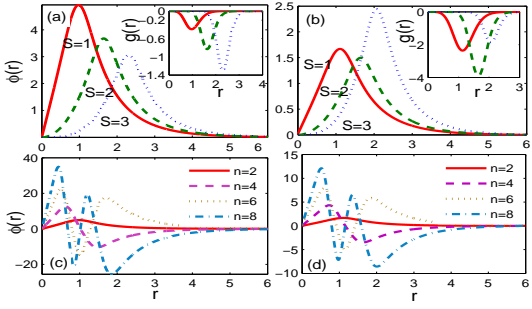}
\caption{(Color online). (a) Exact vortex solitons without the
external potential. Inset, the corresponding profiles of the
attractive-nonlinearity coefficient. (c) Exact vortex solitons of
different radial quantum number $n$ with $S=1$, the respective
nonlinearity-coefficient profile being depicted by the solid line
in the inset of (a). Parameters are $c_{3,4}=-\protect\mu
=-g_{0}=1$. (b) and (d): The same as (a) and (c) when the trapping
potential is present, $V=0.01r^{2}$, with $c_{1,2}=3$.}
\end{figure}

We begin constructing exact VS solutions for the attractive nonlinearity ($%
g_{0}<0$) when $E=0$, so that Eq. (\ref{evolution}) is solvable. With the
harmonic potential $V=kr^{2}$, $\rho $ can be found in terms of the
Whittaker's $\mathrm{M}$ and $\mathrm{W}$ functions \cite{whittaker}: $\rho
(r)=r^{-1}[c_{1}\mathrm{M}({\mu }/{4\sqrt{k}},{|S|}/{2},\sqrt{k}r^{2})+c_{2}%
\mathrm{W}({\mu }/{4\sqrt{k}},{|S|}/{2},\sqrt{k}r^{2})]$, where the
restrictions on $\rho $ require $\mu <\mu _{0}=2(1+|S|)\sqrt{k}$ and $%
c_{1}c_{2}>0$. Without the trap ($k=0$), $\rho $ degenerates to $\rho
(r)=c_{3}I_{S}(\sqrt{-\mu }r)+c_{4}K_{S}(\sqrt{-\mu }r)$, with $\mu <\mu
_{0}=0$, $I_{S}$ and $K_{S}$ being the modified Bessel and Hankel functions,
and the constants satisfying $c_{3}c_{4}>0$. In both cases, one has $\rho
(r)\sim r^{-|S|}$ at $r\rightarrow 0$, hence $g(r)\sim r^{6|S|-2}$ and $%
R(r)\sim r^{2|S|}$ at $r\rightarrow 0$, and $\rho (r)\rightarrow \infty $ as
$r\rightarrow \infty $. Thus, the respective nonlinearity is localized and
bounded, and $R(r)$ is bounded too. To meet boundary conditions $\phi
(0)=\phi (\infty )=0$, an exact solution to Eq. (\ref{final}) is chosen as
\begin{equation}
U(R)=\left( n\eta /\sqrt{-g_{0}}\right) \mathrm{cn}\left( n\eta R-K(1/\sqrt{2%
}),1/\sqrt{2}\right) ,  \label{ellip}
\end{equation}%
where $n=2,4,6,\cdots $ with $n/2$ being the radial quantum number, $\eta \equiv K(\sqrt{2}/2)/R(r=\infty )$, and $K(1/%
\sqrt{2})$ is the complete elliptic integral of the first kind.

It follows from Eq. (\ref{ellip}) that $U(R)\sim R$ at $R\rightarrow 0$,
which implies that the amplitude of the exact VS is $\rho U\sim r^{|S|}$ at $%
r\rightarrow 0$, as it should be. Thus, for given $\mu $, $S$, and
nonlinearity strength $g_{0}$ (and $k$, in the presence of the trap), one
can construct an \emph{infinite number} of exact VSs with $n/2$ bright rings
surrounding the vortex core, as shown in Fig. 1. Although these VSs share
the same chemical potential, their energies increase with the increase of
even number $n$.

\begin{figure}[t]
\includegraphics[width=6.5cm,height=3.5cm]{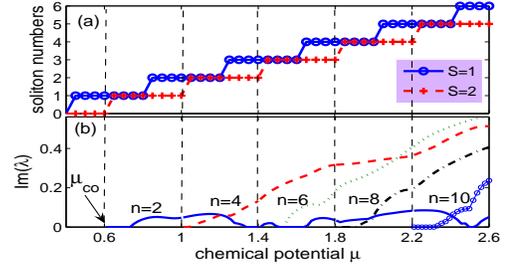}
\caption{(Color online). (a) The number of numerically found vortex-soliton
modes versus the chemical potential in the case of the repulsive
nonlinearity, with the harmonic trap. (b) The largest instability growth
rate for numerically found vortices with $S=2$, $E=1$ and $k=0.01$.}
\end{figure}

Next we consider the repulsive nonlinearity, $g_{0}>0$, and include the
harmonic trap to confine the system. In this case, the existence of
elliptic-function solutions to Eq. (\ref{final}) requires $E>0$, hence Eq. (%
\ref{evolution}) is a nonlinear equation, which can be solved only in a
numerical form. To construct the VSs in this case, one requires $\rho \sim
r^{-|S|}$ (for $S\neq 0$) at $r\rightarrow 0$, hence the nonlinear term in
Eq. (\ref{evolution}) may be neglected near $r=0$. Thus, $\rho$ is similar
to the Neumann function, $Y_{S}(\sqrt{\mu }r)$, at $r\rightarrow 0$ for $\mu
>0$ (for $\mu <0$ it can be checked that VS solutions do not exist). On the
other hand, $\rho \rightarrow \infty $ at $r\rightarrow \infty $, due to the
the presence of the harmonic trap. Further, term $Er^{-2}\rho ^{-3}$ with $%
E>0$ in Eq. (\ref{evolution}) guarantees the sign-definiteness of $\rho (r)$%
. Therefore, taking small $r_{0}$ as an initial point and using the Neumann
function and its derivative at $r=r_{0}$ as initial conditions, one can
numerically integrate Eq. (\ref{evolution}) to obtain $R(r)$ and $g(r)$;
then VSs can be constructed in the numerical form, using the exact solution
to Eq. (\ref{final}),
\begin{equation}
U(R)=\sqrt{{2(E-B^{2})}/{g_{0}}}\mathrm{sn}\left( BR,\sqrt{{E}/{B^{2}}-1}%
\right) ,  \label{ellip_sn}
\end{equation}%
subject to a constraint with even numbers $n$,
\begin{equation}
\sqrt{{E}/{2}}<B\equiv n{K(\sqrt{E/B^{2}-1})}/{R(\infty )}<\sqrt{E}.
\label{B}
\end{equation}%
From Eq. (\ref{B}) it follows that $n<n_{\max }=2R(\infty )\sqrt{E}/\pi $.
This implies that there is a \emph{finite number} of the VS modes (or none,
if $n_{\max }<2$), in contrast to the case of the attractive nonlinearity,
where the infinite set of exact VSs was constructed. Figure 2(a) shows the
number of the numerical VS solutions versus $\mu $, demonstrating that the
cutoff value of the chemical potential is same as for the exact VSs with the
attractive nonlinearity, and the number of VSs jumps at points $\mu =\mu
_{j}^{(S)}\equiv 2(2j+|S|-1)\sqrt{k}$, with $j=1,2,3,\cdots $, which is
precisely the $j$-th energy eigenvalue of the vortex state in the
corresponding linear Schr\"{o}dinger equation. Thus, there are $j$
numerically exact VSs, associated with expression (\ref{ellip_sn}), where $%
n=2,4,6,\cdots ,2j$, in interval $\mu _{j}^{(S)}<\mu \leq \mu _{j+1}^{(S)}$.
A similar result is true for the 1D GPE with the optical-lattice potential,
where $n$ families of fundamental gap solitons exist in the $n$-th bandgap
\cite{wubiao}. Figure 3 displays a characteristic example of the numerically
found VSs, in the case when four of them exist. Similarly, it can be shown
that there are infinitely many VSs, but just the first $j$ modes do not
exist, when $\mu _{j}^{S}\leq \mu <\mu _{j+1}^{S}$ for the attractive
nonlinearity, which demonstrates that the repulsive and attractive
nonlinearities are \emph{mutually complementary} ones, in this respect.

\begin{figure}[t]
\includegraphics[width=8cm,height=2cm]{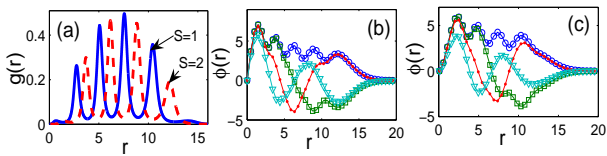}
\caption{(Color online). (a) Modulation profiles of the repulsive
nonlinearity for numerically found vortices, which are displayed in (b) for $%
S=1$ and in (c) for $S=2$, in the presence of the harmonic trap.
Opened circles, squares, solid circles, and triangles denote
solutions with $n=2$, 4,
6, and 8, respectively. Parameters are $E=1$, $\protect\mu =2$, $%
k=g_{0}=0.01 $.}
\end{figure}

Solutions can also be found for 2D fundamental solitons ($S=0$) supported by
the following \textit{two-tier} nonlinearity, with constant $g_{r}$:
\begin{equation}
g(r)=g_{r}~(\mathrm{at}~0\leq r<r_{0}),~{g_{0}}{\rho ^{-6}r^{-2}}\ \left(
\mathrm{at}~r\geq r_{0}\right) .  \label{gg}
\end{equation}

For $r\geq r_{0}$, exact solutions can be found in the same way as above,
except that now $R(r)\equiv \int_{r_{0}}^{r}ds/[s\left( \rho (s)\right)
^{2}] $. At $r<r_{0}$, if $r_{0}$ is small in comparison with the spatial
scale of the external potential (for instance, $r_{0}\ll \sqrt{1/k}$, in the
presence of $V=kr^{2}$), $\phi (r)$ may be approximated by a constant, $\phi
=\sqrt{[\mu -V(0)]/g_{r}}$. Because $\phi (r)$ and $\phi ^{\prime }(r)$ must
be continuous at $r=r_{0}$, one then requires $\rho ^{\prime }(r_{0})=0$ and
$dU(0)/dR=0$, which leads to $g_{r}=[\mu -V(0)]\left[ \rho (r_{0})U(0)\right]
^{-2}$. In this case, the solution to Eq. (\ref{final}) is given by Eqs. (%
\ref{ellip_sn}) and (\ref{B}), with $BR$ replaced by $BR+K(\sqrt{E/B^{2}-1})$%
, and $n=1,3,5,\cdots $ for the repulsive nonlinearity, to make $\phi
(\infty )=0$. Similar solutions can be constructed for the attractive
nonlinearity. Examples of the fundamental solutions are presented in Fig. 4.

\begin{figure}[t]
\includegraphics[width=8cm,height=2.5cm]{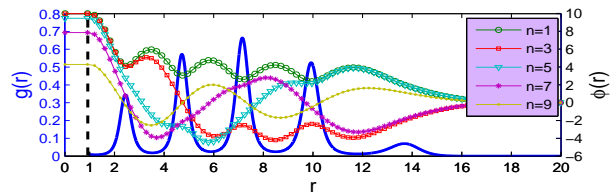}
\caption{(Color online). Fundamental solitons supported by the repulsive
two-tier nonlinearity, see Eq. (\protect\ref{gg}). The constant values of $%
g_{r}$ for $n=1,3,5,7$ and 9 are respectively $0.0200,0.0201,0.0224,0.0325$,
and 0.1093. Parameters are $\protect\rho (r_{0})=1$, $\protect\rho ^{\prime
}(r_{0})=0$, $r_{0}=E=1$, $k=g_{0}=0.01$, $\protect\mu =2$.}
\end{figure}

We employed the linear stability analysis and direct simulations
to verify the stability of the solitons. For the attractive
nonlinearity, we have found that all exact VSs are subject to the
azimuthal modulational instability without the external potential.
As a result, the VSs break up, and eventually collapse. However,
the lowest-order exact VSs with $n=2$ are \emph{stable} in the
presence of the harmonic trap, if $\mu $ is close enough to the
cutoff value $\mu _{0}$, while the amplitude of the VS does not
become small, see Fig. 5. We stress that, in numerous previous
works, a stability region for VSs trapped in the harmonic
potential was found solely for $S=1$, \emph{all} vortices with
$S\geq 2$ being conjectured unstable \cite{BECvortex}. The present
results report the first example of \emph{stable} vortices with
$S\geq 2$ in the trapped self-attractive fields.

%(in agreement with recent works which demonstrate that a purely
%nonlinear trap cannot support stable vortices \cite{2D})

In the case of the repulsive nonlinearity combined with the
harmonic trap, the numerically found VSs can be stable for
\emph{every} $n$ at which they exist, within some region of values
of $\mu $, see Fig. 2(b). This finding is remarkable in the sense
that previous works did not report stable trapped vortices in
higher-order radial states. Moreover, Figs. 2(b) and 6
demonstrates that VSs with the largest number of rings may be
\emph{more stable} than its counterparts with fewer rings. An
explanation to this feature is that local density peaks place
themselves in troughs of the nonlinearity landscape, thus lowering
the system's energy. Similar properties are featured by the
fundamental solitons. Numerical simulations show that those VSs
with $n=2$ which are unstable either split into vortices with
lower topological charges or exhibit a quasi-stability,
periodically breaking and recovering the axial symmetry [Fig.
6(a)], similar to what was previously observed in the case of the
attractive nonlinearity \cite{BECvortex,stable}, while unstable
VSs with $n=4,6,\cdots$ ultimately evolve into vortices located
close to zero-amplitude points.

\begin{figure}[t]
\includegraphics[width=6cm,height=1.7cm]{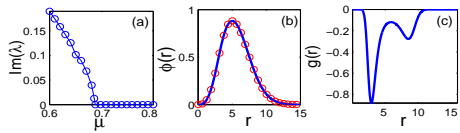}
\caption{(Color online). (a) The largest instability growth rate
for exact vortex solitons with $S=3$ and the attractive
nonlinearity. (b) A stable vortex (solid lines, with circles
showing its profile at $t=200$, when it was initially perturbed by
random noise), and (c) the corresponding nonlinearity coefficient
when $\protect\mu =0.7$. Here $g_{0}=-1000$, $k=0.01$, and
$c_{1,2}=3$. }
\end{figure}

\begin{figure}[t]
\includegraphics[width=6cm,height=1.7cm]{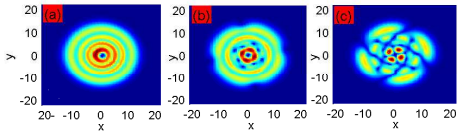}
\caption{(color online). (a) The result of the quasi-stable
evolution of a
numerically found vortex with $n=2$, in the case of the self-repulsion, at $%
t=1700$. (b, c) Unstable vortices with $n=4$ and $6$ at $t=60$ and
$70$, respectively. Parameters are $S=2$, $E=1$, $g_{0}=k=0.01$,
$\mu=1.82$. The vortex with $n=8$ are stable.}
\end{figure}

In conclusion, we have constructed exact solutions of the
Gross-Pitaevskii equation for solitary vortices, and approximate
ones for fundamental solitons in the framework of the GPE/NLSE
with 2D axisymmetric profiles of the modulation of the
nonlinearity coefficient, and harmonic trapping potential. We have
demonstrated that the attractive/repulsive nonlinearity supports
an infinite/finite number of exact VSs. In particular,
\emph{stable} VSs with vorticity $S\geq 2$, as well as those
corresponding to higher-order radial states, have been produced
for the first time. The results suggest a scenario for the
creation of stable vortex solitons in BEC and optics, which have
not been as yet observed in experiments. The necessary BEC
nonlinearity landscape can be built by means of the
Feshbach-resonance technique. The corresponding nonuniform
magnetic field may be created by a micro-fabricated ferromagnetic
structure integrated into the matter-wave chip \cite{exper}, or
one can use the respective pattern of the laser beams. In optics,
the same nonlinearity landscape may created by a nonuniform
distribution of nonlinearity-enhancing dopants. The method
developed in this work for finding the exact solutions,
which is based on reducing the 2D equation to the solvable system of Eqs. (%
\ref{evolution}) and (\ref{final}), can be applied to other models. In
particular, a challenging problem is to devise a physically relevant model
admitting exact 3D solitons.

This work was supported by the NNSF of China (Grants No. 10672147, 10704049
and 10934010), the NKBRSFC (Grant No. 2006CB921400), PNSF of Shanxi (Grant
No. 2007011007), and the German-Israeli Foundation through grant No.
149/2006.

\end{document}